# Security Management Model in Cloud Computing Environment

Seyed Hossein Ahmadpanah[1]

*Abstract*— **In the cloud computing environment, cloud virtual machine (VM) will be more and more the number of virtual machine security and management faced giant Challenge. In order to address security issues cloud computing virtualization environment, this paper presents a virtual machine based on efficient and dynamic deployment VM security management model state migration and scheduling, study of which virtual machine security architecture, based on AHP (Analytic Hierarchy Process) virtual machine deployment and scheduling method, based on CUSUM (Cumulative Sum) DDoS attack detection algorithm, and the above-described method for functional testing and validation.**

## I. Introduction

Cloud computing [1] As a new network computing model based on resource virtualization [2] On the basis of the data center, by consultation, in the form of services provided to users on demand, scalable computing resources to meet the QoS requirements. With the cloud Operators development, virtualization technology in all walks of life more and more attention, leading to more and more users to move their data and applications Move to a virtualized environment, cloud virtual machine (Virtual Machine) number will be more and more. Therefore, the virtual machine Effective deployment and migration to achieve efficient use of physical resources has become a challenge for virtualization management; on the other hand, a malicious user by renting a large number of virtual machines can initiate TCP SYN Flood attacks, the external environment cannot distinguish the attack Cloud those virtual machines, such attacks are subtler, more convenient. etc. [3] Proposed based on virtual mobility State Migration Technology virtual machine cluster scheduling program. Yamuna, etc. [4] We discuss the implementation of dynamic move in KVM virtualized environment Shift mode, analyzes the safety and reliability of dynamic migration exist. Danev, etc. [5] Analysis based vTPM Principles and methods of safe migration of virtual machines to study the method by hardware means to ensure the safety of live migration. But through Points and had in-depth study found that the above method is not enough to guarantee the security of virtualized environments, therefore, based on this paper Efficient deployment and management of virtual machine security model for dynamic migration of one of the key technologies research and implementation.

## II. VIRTUAL MACHINE SECURITY MANAGEMENT MODEL

Figure 1 shows a virtual machine security management model that can be divided into four parts:

1) the multiple physical servers Virtual machine management system;

2) virtual machine condition monitoring system;

3) Based on AHP and live migration of virtual machine technology deployment and scheduling methods;

4) Based on CUSUM algorithm DDoS attack detection mechanisms.

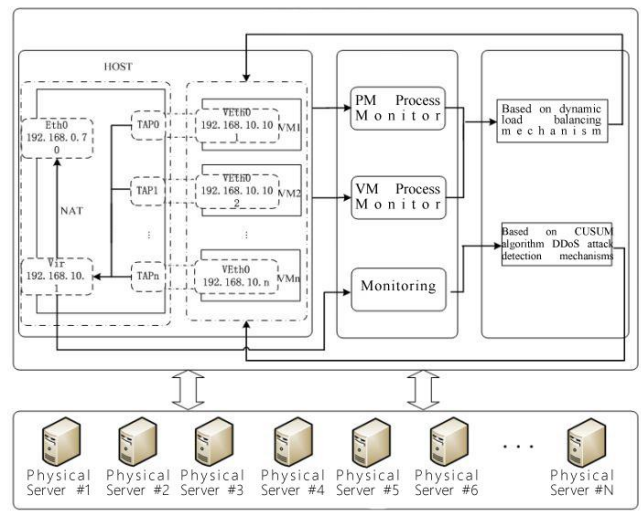

figure1 - virtual machine security management model

Here the key management module VM security model is discussed:

### A. virtual machine management system

To the physical resources of the physical servers through virtualization technologies into a unified abstract, you can dynamically manage logical resource pools, decoupled achieve physical, and users through web application initiates a request to obtain virtual machines cloud computing environment computing resources, so that multiple users can share resources on the same physical server, to avoid idle physical resources Caused by waste, improve resource

1- Seyed Hossein Ahmadpanah is with Department of Computer and Information Technology, Mahdishahr Branch, Islamic Azad University, Mahdishahr, Iran (e-mail: djalicrt@gmail.com).

utilization of physical servers. Meanwhile, in order to achieve the associated virtual machines and cloud storage services, the Department of EC set the following features: access to the web user interfaces, on-demand dynamic application virtual machine, shut down the virtual machine, start the virtual machine, undo the virtual machine, Telnet virtual machines, virtual machine synchronization cloud storage, synchronization and other functions on a virtual machine cloud storage.

*B. migration technology based on dynamic load balancing mechanism*

Virtual machine migration refers to the virtual machine running (source host) on a single host to migrate to another host (destination host) on the run.

Online virtual machine migration (live migration) refers to the whole migration process, the virtual machine pauses (Downtime) is very short, the services running on a virtual machine can always respond to user requests, to ensure that the virtual machine environment transparent to the user. Because users can cloud the virtual machine environment free operation, including the withdrawal of the application and virtual machine pin virtual machine operations. When a user within a short time in the cloud environment at the same time withdrew multiple virtual machines, it will lead to between physical server's load imbalance, a large number of virtual machines running on some physical server tasks, and some of the physical servers are sitting idle, which also will result in a virtual machine deployed on a heavily loaded physical server is not well in the full process of providing services to users QoS requirements. When the existence of the physical server load unevenly, the theoretical nature of resource balancing strategy can be summarized as an optimization, but the property right for each virtual machine weight is not the same, some virtual machines are CPU-intensive, the performance demand is its CPU resource requirements are relatively high; some memory-intensive, the performance requirements of its memory requirements relatively high. Therefore, when the resources of the physical server virtual machine load balancing require a combination of the characteristics of demand for resources, avoid high CPU requirements of a virtual machine is deployed in a relatively scarce CPU resources, memory resources are idle physical servers on the server, this will result in a virtual machine to provide services cannot meet the QoS, which will also result on the same physical server virtual machines competing for CPU resources, thus affecting other virtual machines to provide quality service. In summary, based on AHP virtual machine detailed functional deployment and scheduling method comprising: monitoring statistical characteristics of the physical state of the server, understand the type of virtual machine resources as well as its access characteristics, the characteristics of the virtual machine resource analysis, and evaluation of the physical servers on this basis, to find the most suitable deploy or migrate physical servers to optimize resource allocation virtual machine cluster.

*C. DDoS attack detection method based on CUSUM Algorithm*

In SYN Flood, represented by a distributed denial of service attack using TCP / IP three-way handshake in the presence of unsafe hidden suffering from the attack by multiple attacks originator to the destination host sends a SYN packet to be attacked, but destination is received in SYN + ACK packet is not made after the reaction, the other can attack by source IP address masquerading initiator, the main cause being attacked SYN + ACK packet issued by the machine does not respond, while the destination host team will create a large number of connections for these attacks Started Column, because it did not receive an ACK packet has been the initiator of the attack maintained these connections queue, resulting in a large consumption of resources consumption is not released, leading to some of the normal requests cannot provide services, showing a strong destructive and difficult to prevent, security for the Internet, integrity, availability, and so posed a grave threat. Therefore, based on CUSUM Algorithm DDoS attack detection mechanism main features include: virtual machine network traffic statistics information, including SYN packet and a FIN | RST the number of packets relations; design and implement improved CUSUM algorithm to quickly detect the source; malicious virtual machines deal with.

Wherein the modified CUSUM algorithm is based, from a normal TCP connection establishment to the end, there is a symmetric relationship: single SYN packet and a FIN | RST packet is paired, SYN and FIN + number roughly equal number of RST. When a DDoS attack occurs, SYN and FIN + RST number of packets number of packets in one of the two will be far more than the other too more, by identifying changes in the difference between the two to be detected.

### III. VIRTUAL MACHINE SECURITY MANAGEMENT OF KEY TECHNOLOGIES

*A. AHP-based virtual machine live migration technology*

When running on a small number of virtual machines under the cloud, and circumstances of each physical server under load are low, due to the rising cost of electricity, energy operators face pressure increasing, the cloud should be running a virtual centralized scheduling machines on fewer physical servers, and will be treated no dormant virtual machine tasks that part of the physical servers, to reduce energy consumption. Major companies have actual cases in this regard: Cassatt launched a project called Cassatt Active Response software solutions to achieve physical server automatic sleep and wake conditional, wakeup condition package Including pre-set time (for example, at the end of working day and weekends off the server) and application usage (if you specify Service-level application software reaches the specified limit; the new server will start); IBM introduced a called Active Energy Manager product, which is an upgraded version of IBM Systems Director, an increase of energy control options through Lowering processor clock frequency, or let the server when the processor is idle sleep to improve performance per watt; VMware company in order to achieve the goal of energy conservation in VMware

Infrastructure distributed added energy management entity Now the physical server so that part of the physical servers in the task is not difficult, there is a big surplus of computing resources that are automatically sleep, when the physical server load and run automatically wake of generally too high when sleeping physical server.

When a larger number of virtual machines in the cloud computing environment is running, and each physical server load general bias under the high case, first wake-sleep or physical server downtime through appropriate technology to obtain services from a physical data center the performance monitoring data and virtual machines to physical resource usage, while the virtual machine is defined as a hotspot sensitive resources, and so the virtual machine classification.

### B. Start the virtual machine for the first time

Read by the central dispatch center with data from the virtual machine has the same hot spot to be deployed in a virtual machine running on the matter the use of physical resources, such as the type shown in formula (1) for the M1, to estimate the use of physical resources after the virtual machine starts circumstances.

$$M1=\{CPU1.MEM1.BandWidth1\} \quad (1)$$

At the same time, performance monitoring statistics data of all physical servers, such as formula (2) referred to as M_P.

$$M\_P=\{CPU\_P.MEM\_P.BandWidth\_P\} \quad (2)$$

In order to ensure that after the deployment of the virtual machine does not affect other virtual machines to provide normal service, the server must meet the candidate of formula (3).

$$M\_P + M1 < M\_T \quad (3)$$

M_T which represents the virtual function using a threshold vector physical resources.

Then the AHP method based on the hot virtual machine to assess the weight vector obtained on CPU, memory and bandwidth of the three types of resources, such as formula (5), referred to as Vector.

$$[\blacksquare(CPU@MEM@BandWidth)] = [\blacksquare(W1@W2@W3)] \quad (4)$$

Next, the resulting vector and M1, M_P, according to the formula (4-8) Evaluation of the candidate physical servers, to the result set, denoted by Result.

$$Result = Vector * M\_P \quad (5)$$

Finally, select the candidate physical server has the minimum evaluation value in the result set Result, the virtual machine on which to deploy and start.

### C. virtual machine restarts

Read by the central dispatcher virtual machines to be deployed in the past, during the operation of the physical resources of intelligence from the data center conditions, such as the formula (6) is shown in the M2 recorded, to use the estimate of the physical resources of the virtual machine after the start situation.

$$M2=\{CPU2.MEM2.BandWidth2\} \quad (6)$$

At the same time performance monitoring statistics data of all physical servers M_P.

Similarly, in order to ensure that the physical server has enough free resources available to virtual machines, physical servers candidate must be satisfying equation (3).

Then the AHP method based on the hot virtual machine to assess give weight vector, by scoring vector to vector and M2, M_P, according to equation (5) evaluation of the candidate physical server to obtain the results set Result.

Finally, select the candidate physical server has the minimum evaluation value in the result set Result, the virtual machine on which redirect deployed and started. Since the method used is NFS storage structure of the virtual machine image files stored in the on a common medium, and therefore in the process of deploying virtual machines in a virtual machine image file transfer, synchronization problems can be ignored.

### D. Dynamic migration Virtual Machine

When there is some physical server load is too high and exceeds the predetermined threshold value, and some of the physical servers but in the low load state, affecting QoS operate its virtual machines on the provision of services, then you need on the physical server virtual machine migration, load balancing purposes. First, get the current resources of the physical server from the data center usage, such as the formula (7) referred to as M_P1, at the same time to get the current resource usage of other physical servers, such as formula (8) FIG denoted M_P.

$$M\_P1=\{CPU\_P1.MEM\_P1.BandWidth\_P1\} \quad (7)$$

$$M\_P=\{CPU\_P.MEM\_P.BandWidth\_P\} \quad (8)$$

For all virtual machines running on the physical server, get them running so that the average physical resources with the situation and calculate their weight vector, such as the formula (9) referred to as M3, is used to estimate the migrated virtual machine to a physical resource usage.

$$M3=\{CPU3.MEM3.BandWidth3\} \quad (9)$$

In order to ensure the migration of virtual machines to physical servers does not affect the object after the other virtual machines to provide normal services, the desired product management server need to satisfy equation (10), which represents the threshold M_T vector virtual machine to a physical resource used for M_P Vector remaining physical server and each virtual machine, according to the formula (5) evaluation of all physical servers, the result set referred to as Result. In the same manner, evaluate current physical servers, such as (11), the result is noted as Result '.

$$M\_P + M3 < M\_T \quad (10)$$

$$Result = Vector * (M\_P1 - 3) \quad (11)$$

Result of the result set of the sort the records, and in turn with the new result set Result 'according to the equation (12) for comparison, if the result is less than the current value of the evaluation of physical servers, then take the next set of records, migration of virtual machines to physical server's purposes, until the presence of the collection value smaller than the evaluation value of the evaluation of the current physical server. Otherwise, do not migrate.

$$\min(sort(Result), Result) \quad (11)$$

Because this method uses a storage structure NFS, image files, virtual machines are stored in a common medium, therefore, in the process of deploying virtual machines can ignore the virtual machine image file transfer, synchronization process like a matter of time [9].

*E. DDoS Attack Based on CUSUM Algorithm*

CUSUM (Cumulative Sum) is a method of calculating the cumulative and by analyzing a steady sequence, and detecting the sequence so that the mean square stability and other indicators point to measure the change. CUSUM algorithm DDoS attack detection should be used mainly based on pre-attack SYN and ACK, FIN and other packets in the memory for some time in an approximate steady sequence. For example, when a user initiates a TCP connection, it first sends a SYN packet to the destination node, then the destination node returns a SYN + ACK response data packet, the initiator will return a final ACK packet end times handshake agreement, it is clear that the number and the number of ACK packets of the process of SYN packets in a balanced state, which is an approximate steady sequence. When the sequence is changed, it means the network flooded with SYN or ACK packet does not respond, that the presence of DDoS attacks. [6] Statistical unacknowledged by the number of packets and the total number of packets ratio YAN, forming a time-based statistics sequence, and then improved nonparametric recursive CUSUM (cumulative sum) algorithm [7-8] to quickly detect the source DDoS attacks. While this study focused on the physical network in the attack end passively detected so as to achieve the purpose of defense, there is no ability to identify the originator of the positioning attack, while not be applied to a variety of network topologies combining virtual machine network.

Kernel-based Virtual Machine (KVM) is open in October 2006 by an organization called Qumranet proposed hardware-based virtualization (Intel VT or AMD-v) virtual machine solution that uses host-based VMM model, the entire Linux kernel as a Hypervisor, namely increasing the Linux virtual machine monitor function to provide technical support virtualization extensions can take advantage of the Linux kernel memory management and process scheduling policy, so as to form A lightweight virtual machine monitor. After cutting Qemu and KVM through cooperation, on demand dynamically create a virtual machine, start the virtual machine, shut down the virtual machines and virtual machine operations such revocation. Use KVM can be dynamically created on demand and run multiple virtual machines, however, it is a simple process of Qemu virtual machines on the physical server performance.

In SYN Flood, represented by a distributed denial of service attack using TCP / IP three-way handshake in the presence of unsafe hidden suffering, showing a strong destructive and difficult to prevent, to Internet security, integrity, availability, and so posed a grave threat. Therefore, this paper based on KVM, designed DDoS attack detection method based CUSUM algorithm, Key features include:

1) information about the virtual machine network traffic statistics, including the relationship between the number of SYN packets and FIN + RST packet;

2) design and implement improved CUSUM algorithm to quickly detect the source; malicious virtual machine processing.

Wherein the improvement on the basis of CUSUM algorithm is, in the normal TCP connection from the establishment to the end, there is a pair said relationship: SYN packets and FIN | RST packet is paired, SYN FIN + RST number and the number of substantially equal.

3) When a DDoS attack occurs, the number of the number of SYN packets or FIN + RST packet will be far greater than The other, by recognizing this change, achieve detect attacks.

IV. FUNCTIONAL TEST

Test environment is as follows: three physical servers, Red hat enterprise Linux 4 operating system.

Test tasks: three physical servers a simple cloud computing environments through a unified interface provides virtual machine rental service, users in the cloud computing environment, according to a unified order to apply four virtual machines, each group of five, a total of server, Comparison of the application process and load consumption of resources of each physical server after the application. Because virtual machine application process physical resources and the time that the main bottleneck between physical server load imbalance caused by the process of dynamic equilibrium, where the virtual machine live migration bottleneck is the physical process of dynamic equilibrium of resources and time, therefore, for the consumption of resources can with the number of virtual machines live migration to roughly quantify.

Figure 2 shows the results of a virtual machine environment for virtual machine A hierarchical analysis. A virtual machine by acquiring the resource characteristics calculates the weight vector, then combined resource utilization per physical server based on the weight vector calculated scores for each physical server in the hierarchy analysis. The figure shows that for virtual machine A, its weight vector [0.2,0.6,0.2], the use of resources acquired through a combination of individual physical servers, comprehensive calculated: On the physical server A score of 50.08 in the physical server B score is 36.122, physical server C on a score of 42.288. The smaller physical servers score indicates that the current physical server resources to best meet the remaining running virtual machine A, indicating that the pressure to deploy virtual machine A is, the higher the quality of services provided, so choose a physical server B as the best physical server.

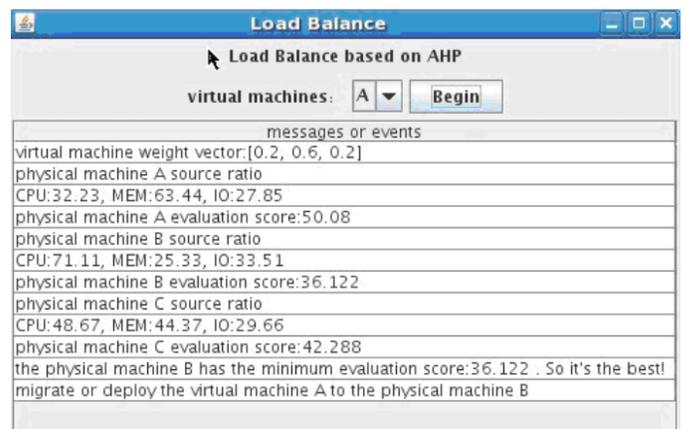

Figure 2: AHP level analysis

Existing studies show that the WAN connection from the SYN packet request issued when FIN or RST packets to terminate the connection issue interval is usually 12 ~ 19s, therefore, before the 10s paper selected for monitoring and statistical sampling interval of SYN packets into the network, and in the 10s of the FIN | RST packet monitor can basically have guaranteed not to miss correspond SYN packet FIN + RST packet. Typically, SYN attack will increase the upper limit of 50% or more, serious and even reached 250% increase; smaller increase for the attack, its strength is not large, causing harm to the network is also smaller. According to a large number of experiments, we chose 1.43 as the threshold for each network mode.

Figure 3 is a KVM virtual machine environment to attack the virtual machine performance monitoring results. This figure shows that the virtual machine at a time, there are 106,242 in the inter-chip connection request SYN and 3 FIN + RST packet is returned, but after the calculated thresholds and no more than 1.43; in the second time slot has generated 107,762 SYN connection requests and 3 FIN | RST packet is returned after the calculated threshold of 1.84, more than 1.43, so we judge the existence of DDoS attacks.

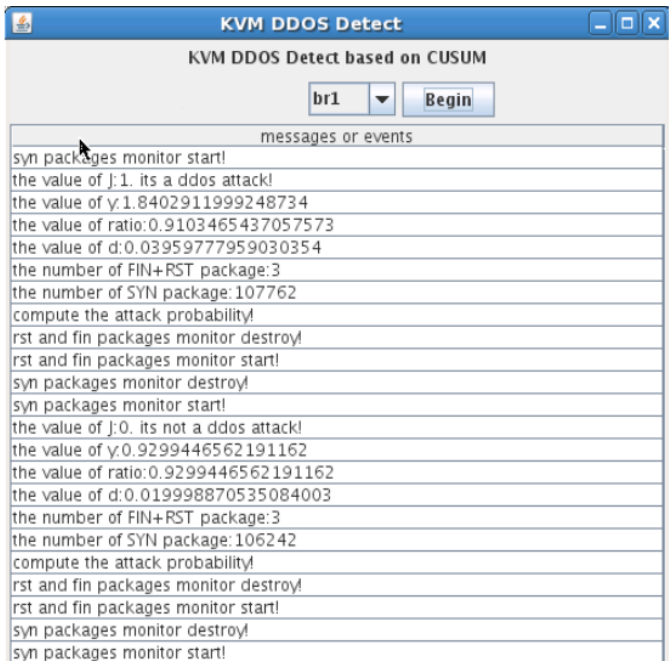

Figure 3 - KVM when attacked Monitoring Results

V. CONCLUSION

Aiming virtual machine security management needs of cloud computing, virtual machine presents a safety management framework model, discussed the functional configuration of the virtual machine management model study in which virtual machines to effectively deploy and dynamic migration mechanism proposed DDoS attack detection method for a based CUSUM algorithm, in time for a malicious user to hire a large number of virtual machines initiated TCP SYN Flood attack detection.